\documentclass[10pt,journal]{IEEEtran}
\usepackage{graphicx}
\usepackage{float}
\usepackage{tikz}
\usetikzlibrary{decorations.pathreplacing}
\usetikzlibrary{shapes,arrows}

\ifCLASSOPTIONcompsoc
  \usepackage[nocompress]{cite}
\else
  \usepackage{cite}
\fi
\ifCLASSINFOpdf
\else
\fi
\begin{document}
%
\title{On the Relevance of Blockchain in Identity Management}
%
%
%
%

\author{Andreas Gr\"uner, Alexander M\"uhle, Christoph Meinel
\IEEEcompsocitemizethanks{\IEEEcompsocthanksitem A. Gr\"uner, A. M\"uhle and C. Meinel are with the 
Hasso-Plattner-Institute (HPI), University of Potsdam, 14482, Potsdam, Germany.\protect\\
E-mail: \{andreas.gruener, alexander.muehle, christoph.meinel\}@hpi.uni-potsdam.de}}

\IEEEtitleabstractindextext{%
\begin{abstract}
The ubiquitous application of emerging blockchain technology in numerous technological projects leads to a tremendous hype. The significantly high prices of digital currencies and initial coin offerings as the new funding approach has fostered the public perception of blockchain as a cure-all and driven the hype even further. In this evolution, a clear view of the reasonable application of blockchain technology is not given and therefore, the purposeful use of traditional technologies is undermined. To clarify this situation, we derive a novel decision model for evaluating the applicability of blockchain technology that considers two key factors: the remediation of central governance and the management of digital objects. Based on these key factors, we closely analyse the domain of identity management for conscious blockchain application. Finally, we examine uPort, Sovrin, and ShoCard as distinct projects in this scope with regard to the inevitable necessity to implement a blockchain by using our decision model.
\end{abstract}

}

\maketitle

\IEEEdisplaynontitleabstractindextext

%
\IEEEpeerreviewmaketitle

\section{Introduction}

\IEEEPARstart{A}{myriad} of projects applies blockchain technology to innovatively address existing challenges \cite{hypeinno2018}. Thus, blockchain technology seems to be perceived by the public as a cure-all to solve numerous issues relating to technology in general or security in particular. Applications range from the irreversible and transparent documentation of financial transactions, peer to peer lending and decentralised exchanges to investments in the financial domain. In identity management, numerous projects are concerned with the creation of self-sovereign identities, new identity provider models and solutions for people lacking official identity documents \cite{blockid2018}. The reason behind why blockchain is often presented as the unavoidable solution in use cases is generally missing. Blockchain technology is a composition of individual parts, for instance, digital signatures, cryptographic hashes and peer to peer communication. Therefore, a dedicated usage of separate components might be more efficient in certain scenarios.\\
The various projects require funding for the development of their ideas until a sufficient revenue stream is established. Using blockchain technology, a new funding and investment approach is created based on the distribution of tokens in exchange for capital. The preliminary token sale is named the initial coin offering. Later on, these tokens are listed on exchanges for trading against other tokens or fiat currencies. The value of a token may fluctuate significantly leading to its speculative nature. Tremendously rising prices of digital currencies at the end of 2017 \cite{williams2017} and tokens, after the initial coin offering, promised a serious return on investment. The overall situation led to blockchain's increased popularity and initiated a certain "gold rush" accompanied by widespread scams \cite{kean2018} and the use of the term blockchain as a buzzword for better marketing and as a way to attract funding.\\
The hype about blockchain technology in general and specifically for identity management goes along with the increase in  projects and funding on the basis of initial coin offering. A detailed rationale for applying blockchain technology is not available for all initiatives, leading to an unclear added value compared to conventional solutions. Taking this as our motivation, we derive a novel decision model for the applicability of blockchain technology that considers the remediation of central governance and the management of digital objects as key factors. Thereafter, we closely analyse the domain of identity management according to the deduced key factors to identify appropriate scenarios. Additionally, we examine uPort, Sovrin and ShoCard, as representatives for distinct implementation strategies of identity projects, regarding the essential use of blockchain technology by applying our novel decision model.\\
The rest of the paper is organized as follows. In Section 2 we outline related work comprising decision models. Key factors for blockchain technology are described in Section 3. A detailed analysis of identity management and blockchain technology is covered in Section 4. Subsequently, we scrutinise the three mentioned projects in Section 5 and present out conclusion in Section 6.

\section{Related Work}

Analysing the effective and necessary use of blockchain technology in various domains lead to the creation of decision models for this purpose. W\"ust and Gervais \cite{wuest2018} propose a flow chart to decide on the usage of a permissionless, public permissioned, private permissioned blockchains or whether to apply conventional technology instead. The decision tree takes into consideration the necessity to store a state, quantity, and identifiability of writers, the possibility to use a trusted third party and the requirement to have public verifiability. In addition to that, the developed decision model is applied in the areas of supply chain management, interbank and international payments, and decentralised autonomous organizations. In contrast, we focus our analysis on the domain identity management and strongly consider the remediation of central governance and the management of digital objects.\\
In 2015 Suchies \cite{suichies2015} published another decision model for the usage of blockchain. The flow chart distinguishes the application of conventional technology, public, private or hybrid blockchains. Requirements relating to a database, shared write access, trust in writers, the need for a trusted third party and confidentiality of transaction data is mainly considered in the decision process. In contrast to our decision model, the nature of the blockchain managed object is not considered. Additionally, the decision process is not applied to a domain.\\
Birch et al. \cite{birch2016} publish the Birch-Brown-Parulava model to decide on the use of either private or public blockchains. The model focuses on the financial sector and takes the communication and integrity of transactions into consideration. In contrast, we focus on the identity management domain and apply the remediation of central governance and the texture of the managed object in our decision process.\\
Lewis \cite{lewis} proposes a decision model that takes a business perspective and considers the general hype about blockchain. The decision process separates the use of blockchain and the application of a regular database. The work flow considers the solution of a real business problem and scrutinises alternatives prior to the invention of blockchain technology. Furthermore, the model incorporates aspects of control, replacement of systems and visibility of data. Compared to our decision model, a dedicated application of the work flow to a certain domain is not available.\\
A further model is published by the Department of Homeland Security \cite{dhs}. This decision model concentrates on evaluating the general use of blockchain. A shared data store, data contributors, modification of data records, the critical nature of the stored data and other factors are taken into consideration by the decision tree. In contrast to our work, a detailed analysis of the identity management domain is not comprised.\\
Overall, several decision models regarding the reasonable use of blockchain technology, covering the types private, public and hybrid, are proposed. The models are applied to the supply chain management, financial technology and decentralised autonomous organization. To the best of our knowledge, the identity management domain is not examined so far.

\section{Key Factors for Applying Blockchain}

Two key factors for evaluating the essential use of blockchain technology are derivable from its origin as decentralised digital cash scheme. These key considerations build the foundation of our decision model.
\subsection{Remediation of Central Governance}
Digital cash schemes are proposed prior to Bitcoin. The electronic cash of Chaum et al. \cite{chaum1990} is an example. However, Nakamoto's Bitcoin proposal solved for the first time the double spending problem without requiring a trusted third party and achieved an entirely decentralised digital cash scheme. In earlier approaches, preventing double spending required a trusted third party as central governance mechanism due to the arbitrarily copyable nature of digital cash \cite{feuer2012}. In Bitcoin, the central governance is remediated by a peer-to-peer consensus algorithm and enforced decentralisation in the blockchain network. Therefore, a key factor for applying blockchain is the remediation of the need to trust a central governance body and distributing trust into a decentralised network of equitable peers without hierarchy.\\
In an alleviated version, trust requirements are transitioned from one party to several parties, potentially not covering all nodes of the blockchain network. To enable the trust shift, the rules that are supervised by the central governance body are required to be digitally represented and enforceable in the decentralised network.\\
Overall, using blockchain technology and introducing centralised trust dependencies counteract each other. Architectures that are built in this regard may better use conventional technology.
\subsection{Management of Digital Objects}
Another key factor for the reasonable use of blockchain is the type of commodity that is managed using a decentralised network. An inherently digital item, comparable to digital cash, is primarily suited to this scenario. There is no need to construct a digital representation of a physical object and link both securely together.\\ 
Additionally, it is important to consider the transition from the outside world into the blockchain environment and vice versa. Digital cash, as a general purpose good, is able to ideally solve this challenge. Coins of digital cash are directly created during mining in the blockchain network. In addition to that, digital cash can be exchanged from and to fiat currencies by using central or decentralised exchanges. Buying goods and services is another channel to transition from the blockchain environment.\\
At the same time, the creation of a digital representation needs to be thoroughly studied to provide the required benefit and support an appropriate use of blockchain. Storing a cryptographic hash of a document on the blockchain, to show that it exists at a specific point in time, is an appropriate scenario for creating a certain association. In contrast, preserving the ownership of valuable art by logging the possessor and characteristics of a picture on a blockchain tends to be ineffective, due to the loose binding between data on the blockchain and the picture as a physical object in the real world \cite{eden2018}.\\
A solution to process external input on blockchain-based applications is an oracle. An oracle is queried during the execution of a decentralised program and provides the required information. Depending on the structure of the oracle, a trusted third party might be introduced and the first key factor is impaired. As an example, SchellingCoins \cite{buterin2014} are a proposal for a decentralised oracle.\\ 
Overall, an artificially created digital association may undermine the original purpose or limit or even eliminate, respectively, the benefit of applying blockchain technology.
\section{Decision Model}
Considering the key factors and further properties of blockchain technology, we devise a new decision model for applying blockchain technology. The decision work flow is shown in Figure \ref{model}. In general, the graph is divided into two sections. These stages refer to the use case for implementing blockchain and the architecture of the actual solution.\\
In the use case section, the first decision step targets the replacement of a trusted third party. In case the objective of the evaluated scenario is not to replace a trusted third party, and therefore does not share trust in a decentralised network of nodes, other technologies are better suited. The replacement of a trusted third party is the primary application of blockchain technology on the basis of decentralised consensus among equitable peers. Subsequently, the texture of the objects that should be managed by the blockchain network is analysed. Either the objects are inherently digital, for instance, a digital currency or a digital representation of the object is possible. The mapping of a physical object to its digital representation needs to avoid the introduction of a new centralised entity that is responsible for the relation. This would again introduce a trusted third party that counteracts the remediation of central governance. Afterwards, the rules, that needs to be enforced on the managed objects, are assessed. In general, a central authority put rules into effect to control certain behaviour. These rules must be digitally enforceable to be distributed in a decentralised blockchain network. A potential source of centralisation is introduced, if rules are required to be exclusively or partially enforced outside the blockchain environment. 
The use case qualifies for applying blockchain technology if all decision steps are appropriately confirmed.\\
The second part of the decision model evaluates the architecture of the solution. First of all, a decision about the demand for a consented and decentralised data storage is required. Data within this storage is decentrally distributed to the nodes and agreed by the nodes of the network. On the one hand, exclusively peer to peer communication should be  considered in case such a data storage is not required. On the other hand, if a data storage is required, the next step is the analysis of special permissions for the actors within the solution architecture. Actors are reflected by the nodes of the blockchain network itself or any other additional entity that interacts with the blockchain-based architecture. No special privileges are required by the nodes in case an unpermissioned blockchain is used. Subsequently, the end of the work flow is reached, determining a suitable architecture for implementing blockchain technology. In case a permissioned blockchain is applied that requires special rights or differentiates the rights of the nodes, an analysis on how these privileges are granted is conducted. If the privileges are assigned by a trusted third party, a centralised dependency is introduced and the architecture may disqualify itself for using blockchain technology. Otherwise, the application of blockchain seems to be reasonable when privileges are granted in a decentralised way and not by a trusted third party.
\section{Applicability of Blockchain in Identity Management}
In this section we examine the identity management domain outlining the functional scope, presenting trusted third parties as well as managed artefacts. We evaluate the general applicability of a blockchain concerning the first stage of our decision model. 
\subsection{Identity Management}
Identity management subsumes the administration, in terms of creation, modification, and deletion, of digital identities. A digital identity is the electronic representation of a person or entity in the real world. The digital identity comprises attributes and is referenced by an identifier \cite{williamson2009}. Attributes are used to specify information about identity.\\
Digital identities are used in the identification, authentication and authorization process to legitimately grant access in applications for the respective subject. Identification describes the process to reference a certain entity by an identifier. Authentication is the operation to ascertain that an entity is in control of a specific digital identity. The owner proves control by using a credential. Authorization is the process to validate that a digital identity is entitled to execute a specific function or access particular data.

\begin{figure}[H]
\centering
\includegraphics[scale=0.9]{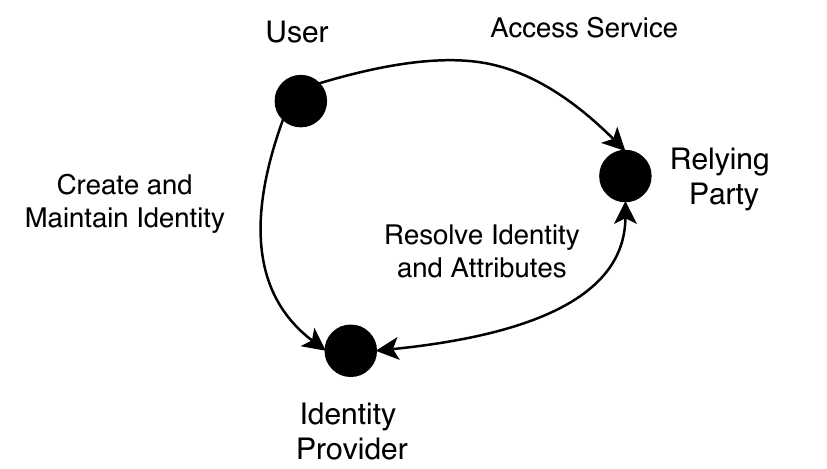}
\caption{Trusted Third Parties}
\label{ttp}
\end{figure}

\subsection{Trusted Third Parties}
Digital identities are provided respectively and used by different actors (see Figure \ref{ttp}). The user or subject creates a digital identity at an identity provider. Additionally, the user provides information about attributes and maintains them to cater for updates. The identity provider offers identity related services, for instance, authentication and authorization. The identity provider may verify the correctness and validity of attributes of a digital identity. Digital identities are used at relying parties. A relying party is a service provider that needs to identify its users and, therefore, utilises the service of an identity provider. A user controls the respective digital identity with a credential. The described entities are classes of actors. For each category, several actors exist and these are commonly used in realistic scenarios. In a specific trust context, the relying party and the identity provider might belong to the same organization. However, in the following analysis, we consider the user, the identity provider and the relying party as distinct trusted third parties for the respective other entities.\\

\textbf{TTP 1: The User} \\
Usually, the user is not seen as a particular trusted third party, but represents a group that is trusted for common factors by the identity provider and the relying party. The user is primarily trusted to keep the credential for controlling the digital identity secret \cite{josang2005}. Thus, the identity can be solely used by the user itself. In case the credentials are published or available to different users, actions that are done with the digital identity cannot be attributed to a specific user. Furthermore, services provided by the relying party to certain users might be consumed by others using the same digital identity. Besides the confidentiality of the credential, the user is secondarily trusted to a particular extent for maintaining that non-essential attributes at the identity provider to be valid and up-to-date.\\

\textbf{TTP 2: The Identity Provider} \\ 
The user and the relying party trust the identity provider to solely allow the legitimate subject to control the respective digital identity and, therefore, to conduct correct authentication \cite{kylau2009}. Additionally, the identity provider is trusted to adhere to agreed data privacy principles. The user expects that its data is kept confidential and is only disclosed to authorised parties when necessary \cite{josang2005}.\\ 
The user and the relying party expect a stable service by the identity provider according to contractual agreements. Especially, the user requires no unfounded rejection on using the service that is offered by the identity provider or arbitrary revocation of the owned identity. Furthermore, the relying party trusts the identity provider for the validity and correctness of critical attributes. Usually, the identity provider is mandated to verify attributes that are essential for the relying party.\\

\textbf{TTP 3: The Relying Party} \\
The user trusts the relying party to accept the digital identity and required attributes stored by the identity provider to access the requested service. The relying party is trusted to provide the offered service and uses the gathered information about the user solely for the consented case. A misuse of user data is neither expected by the identity provider nor the user.

\tikzstyle{decision} = [diamond, draw, text width=3.5em, text badly centered, node distance=3cm, inner sep=0pt]
\tikzstyle{line} = [draw, -latex']
\tikzstyle{cloud} = [draw, ellipse, node distance=3cm, minimum height=1.5em]
    
\begin{figure*}
\centering
\begin{tikzpicture}
    \node [cloud] (start) at (5,100) {Start};
    \node [decision, label=above right:TTP to be replaced?] (ttp) at (5,97) {TTP};
    \node [decision, label=above right:Digital object to manage?] (digitalgood) at (5,94) {Digital Object};
    \node [decision, label=above right:Digital mapping possible?] (digitalrep) at (10,94) {Digital Map.};
    \node [decision, label=above right:Centralisation introduced?] (central) at (10,91) {Central.};
    \node [decision, label=above right:All rules digitally enforceable?] (rule) at (5,88) {Rules};
    \node [decision, label=above right:Consented decentralised data storage required?] (storage) at (5,85) {Storage};
	\node [decision, label={[align=left]above right:Actors require special\\ privileges?}] (privilege) at (5,82) {Privilege};    
    \node [decision, label=above right:Privileges granted by TTP?] (pttp) at (10,82) {Privilege TTP};
    \node [cloud] (blockchain) at (5,79) {Blockchain};
    \node [cloud] (noblockchain) at (15,79) {No Blockchain};
    \path [line] (start) -- (ttp);
    \path [line] (ttp) -- node[midway, left] {Yes} (digitalgood);
    \path [line] (ttp) -- node[midway, below] {No} (15,97)  -- (noblockchain);
    \path [line] (rule) -- node[midway, left] {Yes} (storage);
    \path [line] (rule) -- node[midway, below] {No} (15,88);
    \path [line] (digitalgood) -- node[midway, below] {No} (digitalrep);
	\path [line] (digitalgood) -- node[midway, left] {Yes} (rule);    
    \path [line] (digitalrep) -- node[midway, left] {Yes} (central); 
    \path [line] (digitalrep) -- node[midway, below] {No} (15,94);
    \path [line] (central) -- node[midway, below] {No} (5,91);    
    \path [line] (central) -- node[midway, below] {Yes} (15,91);  
    \path [line] (storage) -- node[midway, left] {Yes}(privilege);
    \path [line] (storage) -- node[midway, below] {No}(15,85);
    \path [line] (privilege) -- node[midway, below] {Yes}(pttp);
    \path [line] (privilege) -- node[midway, left] {No}(blockchain);
    \path [line] (pttp) -- node[midway, below] {Yes}(15,82);
    \path [line] (pttp) -- node[midway, left] {No} (10,79) -- (blockchain);  
    \draw[decorate, decoration={brace,mirror}] (3.5,98.5) -- (3.5,86.8) node[midway, left, xshift=-0.2cm] {Use Case };
    \draw[decorate, decoration={brace,mirror}] (3.5,86.2) -- (3.5,78) node[midway, left, xshift=-0.2cm] {Architecture };
\end{tikzpicture}
\caption{Decision Model}
\label{model}
\end{figure*}
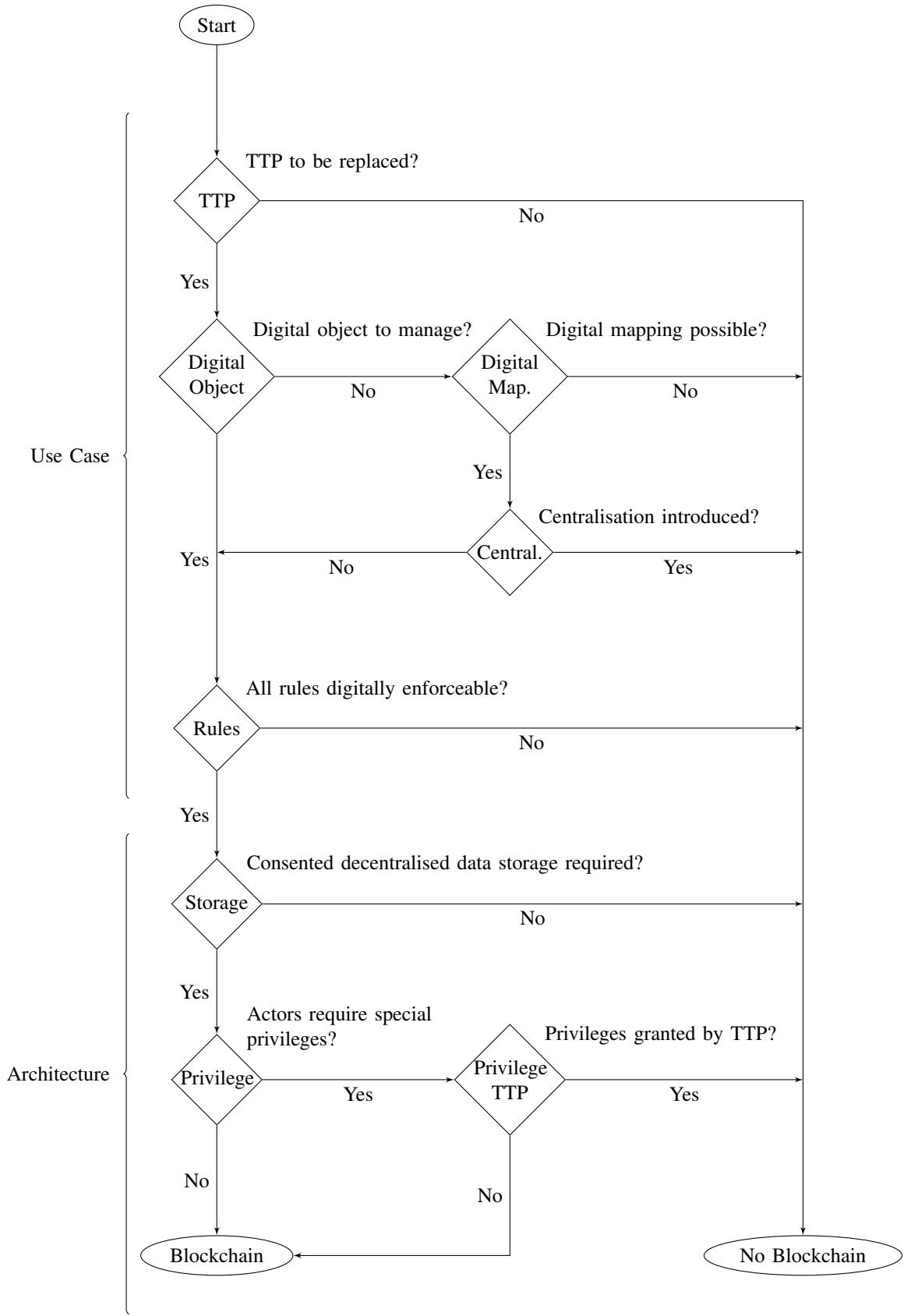

\subsection{Manageable Objects}
In identity management, the administrated objects are digital identities and their associated attributes. Attributes can reflect arbitrary information about the digital identity. We differentiate between two categories of attributes: property and permission. A property is a statement about the digital identity to more closely specify its appearance. This type of attribute relates to information about the owner. A permission is a privilege within an application, for instance, to execute a specific function or to read or write certain data.
\subsection{Evaluation}
We apply the first section of our decision model to generally evaluate the domain of identity management. The replacement of a trusted third party, the management of digital objects and the digital enforcement of rules are considered.\\
The user, the relying party and the identity provider are the essential trusted third parties in identity management. The user is predominantly trusted to keep the authentication credential secret. Blockchain technology does not provide a means to remove this trust requirement. Therefore, the user must still be trusted to keep the credential secret.\\ 
The trust in the relying party targets its behaviour of accepting digital identities, providing an agreed service and using data appropriately. Processes of the relying party can be implemented by using smart contracts on a public blockchain. The processes and its outcome are commonly verifiable as a result. The reduction of trust significantly depends on the implemented processes of the relying party's organizational and business model. External dependencies that act as points of centralisation might be required to fully reflect business processes. Additionally, the processes may differ considerably between the different actors in this category. Therefore, a general assessment of the applicability of blockchain is not possible and we do not continue considering the relying party in the evaluation process.\\ 
The identity provider holds the most trust requirements, from both the user and the relying party, referencing the adequate user authentication, stable service delivery, adherence to data privacy and the verification of attributes. Additionally, different actors in this category have a largely congruent set of processes. Blockchain technology enables a decentralised implementation of an identity provider in the form of smart contracts or as a separate blockchain to remediate the trust requirements. The public verifiability of the implemented processes achieves transparent behaviour of users and relying parties. The general availability of the service is bound in a decentralised way to the participants of the blockchain network. Therefore, the identity provider as core actor in identity management is an ideal trusted third party to be replaced by blockchain technology.\\
The identity provider manages digital identities that are comprised of an identifier and attributes. The identifier and permissions are inherently digital considering the pure digital use and origin in applications. A user's digital identity is referenced by an identifier and permissions are assigned and checked within applications. Therefore, no mapping to the physical world needs to be applied. In the same way, properties are digitally stored and distributed to the respective entities. However, assessing the validity and correctness of a property is challenging, as it reflects a physical characteristic of the user. The age or address of a person, for instance, can hardly be verified without knowing the person. To avoid the introduction of centralisation by relying on a specific verification provider, properties can be modelled as verifiable claims \cite{sporny2018}, that are comprised of claims and attestations. A claim is a statement about the digital identity and an attestation is the verification of this statement. Having numerous attestation issuers that attest one claim several times prevents centralisation towards one verifier. In addition, it might be an approach to establish confidence in a property in a decentralised manner.\\
After devising a trusted third party and the managed objects, the next step is to assess that all rules are digitally enforceable. The identity provider's rules are related to identification, authentication and authorization. These processes are digitally depictable. Therefore, the query for digitally enforceable rules is answered positively.\\
Overall, the identity provider is a trusted third party that applies digitally enforceable rules to digitally manageable objects leading to a scenario where blockchain technology can be applied reasonably. 

\section{Project Review}
There are a large number of blockchain-based identity management projects that target the implementation of a decentralised identity provider. We have chosen uPort, Sovrin, and ShoCard, which provide a wide range of solution architectures to analyse the applicability of blockchain technology according to the second phase of our decision model.
\subsection{uPort}
On the public and unpermissioned Ethereum blockchain, uPort \cite{lundkvist2017} implements a decentralised identity provider using several smart contracts. The proxy, controller and recovery contract principally outlines a digital identity. The proxy contract's address serves as a constant identifier of the digital identity. The digital identity is managed by the controller contract, and the recovery contract provides the means to restore the authentication credential. Furthermore, a central smart contract is used as a verifiable claims registry.\\
A distributed data storage is implicitly utilised on the Ethereum blockchain by taking the proxy contract's address as an identifier. Additionally, the verifiable claims registry directly stores data regarding attributes on the blockchain. Therefore, a consented and decentralised data storage is applied for this solution. As the Ethereum blockchain is public and unpermissioned there are no special privileges required for the nodes of the network. Additionally, there are no other access restrictions introduced that are bound to a trusted third party. Ethereum and uPort are transparently and openly available to the public. In summary, uPort is a decentralised identity provider applying a solution architecture that is a perfect fit for using blockchain technology.
\subsection{Sovrin}
Sovrin \cite{reed2016} is a decentralised identity provider that is implemented as a set of dedicated blockchains for identity management. These blockchains store identity as well as config related information and apply the public and permissioned model. Furthermore, a voting ledger is used to assigned permissions to the nodes. Nodes of the network are differentiated into validators and observers. Validator nodes have the privilege of forging the next block, which includes transactions. In contrast, observer nodes exclusively read the blockchain data. A digital identity consists of an identifier and attributes that are represented as claims and attestations. Sovrin is governed by a foundation adhering to a comprehensive trust framework. The main decision council is the Board of Trustees \cite{reed2017}. Members of this board can appoint new members and select stewards according to rules. Stewards are entities that operate validator nodes.\\
The Sovrin system utilises a blockchain-based data storage to store the identifier and, in certain cases, the attributes on the blockchain. Therefore, the requirement of a consented and decentralised data storage is fulfilled. As Sovrin implements a public and permissioned blockchain, nodes, in particular validators, require special privileges to participate in the network. These privileges are assigned based on a quorum of the Board of Trustees. Depending on the structure of this panel, it can be perceived as a trusted third party that finally makes the decision. In case the Board of Trustees is comprised of a large number of members with a very diverse background, it may not be perceived as a trusted third party.\\
Overall, Sovrin provides a blockchain-based identity management system with the objective to be decentralised. Nevertheless, centralisation is introduced to a certain extent by establishing a strict trust framework and governance bodies. A future diversification of the Board of Trustees will show the evolution to or away from a trusted third party. Therefore, at this time we cannot make a final decision on the applicability of blockchain.
\subsection{ShoCard}
ShoCard \cite{shocard2017} is a blockchain-based identity provider platform that is roughly described in a white paper. Detailed information is, however, lacking. The platform is split into three tiers: the application layer, ShoCard core services, and the blockchain layer. On the application layer, end user and server side solutions exist to connect to the core services tier. The ShoCard core services act as middleware and are comprised of ShoServer, ShoStore, side chains and a caching infrastructure. ShoServer and ShoStore act as an intermediary to negotiate write actions to the side chains and the blockchain layer. The blockchain layer consists of an adaptor to enable the use of different public or private blockchains.\\
ShoCard uses distributed data stores in the form of side chains for identity information. Furthermore, verification data can be stored on public or private blockchains. From the white paper \cite{shocard2017} it can be derived that the service layer tends to be a centralised access point, despite being differently stated.\\
In summary, the proposed blockchain-based identity provider infrastructure seems to implement a significant point of centralisation. Based on this, ShoCard becomes a trusted third party and, from our point of view, the application of blockchain technology might not be mandatory.
\section{Conclusion}
Blockchain technology has been undergoing tremendous hype based on numerous projects and initial coin offerings. A reasonable application needs to consider the remediation of central governance and the management of digital objects as key factors. We incorporated these key factors into a new two-phase decision model. Subsequently, we used the first stage to evaluate the identity management domain in general. The identity provider is an ideal, trusted third party that can be reasonably replaced by decentralising trust using a blockchain. A challenge is to ensure the validity and correctness of properties of a digital identity without introducing dependencies on other trusted third parties. Afterwards, we applied the second phase to diverse projects in this domain and were able to draw specific conclusions. The architecture of uPort ideally utilises the benefits of a blockchain. Sovrin may introduce trust centralisation based on a strict governance framework. The solution architecture of ShoCard may not require a blockchain. 

\ifCLASSOPTIONcaptionsoff
  \newpage
\fi



\bibliographystyle{IEEEtran}
\bibliography{IEEEabrv,bibliography}
%

%








\end{document}